\documentclass[twocolumn,prl,showpacs]{revtex4}
\usepackage{graphicx}
\begin{document}
\newcommand{\re}{\mathop{\mathrm{Re}}}
\newcommand{\be}{\begin{equation}}
\newcommand{\ee}{\end{equation}}
\newcommand{\bea}{\begin{eqnarray}}
\newcommand{\eea}{\end{eqnarray}}
\title{Barotropic index $w$-singularities in cosmology}
\author{Mariusz P. D\c{a}browski}
\email{mpdabfz@wmf.univ.szczecin.pl}
\affiliation{\it Institute of Physics, University of Szczecin, Wielkopolska 15, 70-451 Szczecin, Poland}
\author{Tomasz Denkiewicz}
\email{atomekd@wmf.univ.szczecin.pl}
\affiliation{\it Institute of Physics, University of Szczecin, Wielkopolska 15, 70-451 Szczecin, Poland}
\date{\today}
\begin{abstract}
We find an explicit cosmological model which allows a special type of cosmological singularity
which we call a $w$-singularity. This singularity has the scale factor finite, the energy density and pressure
vanishing, and the only singular behaviour appears
in a time-dependent barotropic index $w(t)$. It is different from
the type IV cosmological singularity in that it does not exhibit
the divergence of the higher derivatives of the Hubble parameter and from the big-brake since
it does not fulfill the anti-Chaplygin gas equation of state. We also find an interesting duality between the $w$-singularities and the big-bang singularities. Physical examples of $w$-singularities appear in $f(R)$, scalar field and brane cosmologies.
\end{abstract}
\maketitle
There has been a strong evidence for the accelerated expansion
of the universe \cite{superrecent} and, in particular, for its superexpansion in the sense of
admitting a very large negative pressure possibly driven by phantom \cite{phantom}
with its exotic evolution towards a big-rip singularity. It opened interest in
investigating more examples of the non-conventional types of matter sources in the universe and exotic singularities accompanying them.

Within the series of these exotic singularities there are the sudden future singularities \cite{barrow04} which
are the singularities of pressure with the finite scale factor and the energy density (also called big-brake \cite{kamenshchik},
when they admit an anti-Chaplygin equation of state), the generalized sudden pressure
singularities \cite{barrow042} which are singularities of the pressure derivatives, and the singularities
of type III and IV according to the classification of Ref. \cite{nojiri}. Type III is a singularity of the pressure and the energy density
with the finite value of the scale factor. The singularity of type IV is a singularity of the higher derivatives of the Hubble parameter
as well as of the time-dependent barotropic index in the barotropic form of an equation of state. However, it allows a finite  value of the scale factor as well as both the energy density and the pressure vanishing.

It is interesting that most of the exotic singularities (except big-rip and type III) are of a weak type, i.e., there is no geodesic incompletness
and the cosmic evolution may eventually be
extended beyond them \cite{lazkoz,adam,celine}. For example, the only physical
characteristic of the sudden future singularities is a momentarily infinite peak
of the tidal forces in the universe. In the case of generalized sudden future singularities this
peak may also appear in the time derivatives of the tidal forces. It has been checked that sudden future singularities
arise in both homogeneous \cite{barrow042} and inhomogeneous \cite{sfs1}
models of the universe. However, the strongest case for their generality is that they plague the cosmological models
based on the loop quantum gravity \cite{wandsLQC} (cf. also \cite{singh09} for investigations of other types of singularities in this
context). More properties of these singularities with respect to quantum theory have also been investigated recently \cite{quantumprod,sfsstabil}.

It is also interesting that exotic type singularity models both may and may not violate the energy conditions. For example,
the phantom models violate the null energy condition, the sudden future singularity models violate the dominant energy condition, while the
generalized sudden future singularity models do not violate energy conditions at all. It then seems that the standard energy conditions do
not work for these exotic types of singularities and the need to formulate a new set of energy conditions appears \cite{marplb05}.
This may also be related to the statefinder diagnostics of cosmology \cite{marplb05,statefin}.
In fact, the sudden future singularities have been verified observationally against supernovae \cite{sfs1}. Amazingly, it emerged that they were possible to appear just in 8.7 Myr in the future. Some other types of exotic singularities, including restrictions on their appearance related to energy conditions violation, were also tested observationally \cite{salvatore05,santos07,gergely09}.

In this paper we suggest yet another type of the exotic singularities which seem to be most similar in nature to the
type IV, as suggested in Ref. \cite{nojiri}. However, we claim that they are different from type IV since they do not lead to any divergence of higher-derivatives of the Hubble function.

Our starting point will be the standard system of the flat geometry Einstein-Friedmann equations
\bea \label{rho} \varrho &=& \frac{3}{8\pi G}
\frac{\dot{a}^2}{a^2}~,\\
\label{p} p &=& - \frac{c^2}{8\pi G} \left(2 \frac{\ddot{a}}{a} +
\frac{\dot{a}^2}{a^2} \right)~,
\eea
with the energy-momentum conservation law
\be \label{conser} \dot{\varrho} = - 3 \frac{\dot{a}}{a}
\left(\varrho + \frac{p}{c^2} \right)~.
\ee
Here $a(t)$ is
the scale factor, $G$ is the gravitational constant, $c$ the velocity of light, $\varrho$ the
energy density, $p$ the pressure. We choose the following form of the scale factor
\be
\label{SF1}
a(t)=A+B\left(\frac{t}{t_s}\right)^{\frac{2}{3\gamma}}+C\left(D-\frac{t}{t_s}\right)^n.
\ee
It contains seven arbitrary constants: $A$, $B$, $C$, $D$, $\gamma$, $n$, and $t_s$. The last of the constants $t_s$ is the time when we expect
the singularity. On the ohter hand, the constant $\gamma$ is an analogous of the constant in the barotropic equation of state $p=(\gamma - 1)\varrho$ in the standard Friedmann model. Having the scale factor (\ref{SF1}), we impose the following conditions
\be
\label{cond1}
a(0)=0, \hspace{0.1cm} a(t_s)=const.\equiv a_s, \hspace{0.1cm} \dot{a}(t_s)=0, \hspace{0.1cm} \ddot{a}(t_s)=0~.
\ee
The first of the conditions (\ref{cond1}) is chosen in order for the evolution to begin with a standard big-bang singularity at
$t=0$ (note that in order to have a big-rip, one would have to impose $a(0) = \infty$, which is equivalent to taking $\gamma<0$).
From (\ref{rho}) and (\ref{p}), it is easy to see that after introducing (\ref{cond1}), the energy density and the pressure
vanish at $t=t_s$. Now we note that the derivatives of the scale factor (\ref{SF1}) are
\bea
\dot{a}(t)&=&\frac{2}{3\gamma}\frac{B}{t_s}\left(\frac{t}{t_s}\right)^{\frac{2}{3\gamma}-1}-\frac{Cn}{t_s}\left(D-\frac{t}{t_s}\right)^{n-1}~,\\
\ddot{a}(t)&=&\frac{2}{3\gamma}\left(\frac{2}{3\gamma}-1\right)\frac{B}{t_s^2}\left(\frac{t}{t_s}\right)^{\frac{2}{3\gamma}-2}\\&+& \frac{Cn(n-1)}{t_s^2}\left(D-\frac{t}{t_s}\right)^{n-2}~,\\
\dddot{a}(t)&=&\frac{2}{3\gamma}\left(\frac{2}{3\gamma}-1\right)\left(\frac{2}{3\gamma}-2\right)\frac{B}{t_s^3}\left(\frac{t}{t_s}\right)^{\frac{2}{3\gamma}-3}~,\\&-&\frac{Cn(n-1)(n-2)}{t_s^3}\left(D-\frac{t}{t_s}\right)^{n-3}~.
\eea
Imposing the conditions (\ref{SF1}), we can
determine the constants $A$, $B$, $C$, and $D$ to be
\bea
\label{A}
A&=&\frac{a_s}{1-\frac{3\gamma}{2}\left(\frac{n-1}{n-\frac{2}{3\gamma}}\right)^{n-1}}~,\\
B&=&n\frac{1-\frac{2}{3 \gamma}}{n-\frac{2}{3 \gamma}}\frac{a_s}{1-\frac{2}{3\gamma}\left(\frac{n-\frac{2}{3\gamma}}{n-1}\right)^{n-1}}~,\\
C&=&\frac{1-\frac{2}{3 \gamma}}{n-\frac{2}{3 \gamma}}\frac{a_s}{\frac{3\gamma}{2}\left(\frac{n-1}{1-\frac{2}{3\gamma}}\right)^{n-1}    -\left(\frac{n-\frac{2}{3\gamma}}{1-\frac{2}{3\gamma}}\right)^{n-1}}~,\\
\label{D}
D&=&\frac{n-\frac{2}{3\gamma}}{1-\frac{2}{3\gamma}}, \hspace{0.2cm} D-1=\frac{n-1}{1-\frac{2}{3\gamma}}~.
\eea
Using the formulas (\ref{A})-(\ref{D}), we have the final form of the scale factor given by
\bea
\label{SFF}
a(t)&=&\frac{a_s}{1-\frac{3\gamma}{2}\left(\frac{n-1}{n-\frac{2}{3\gamma}}\right)^{n-1}}\nonumber
\\&+&\frac{1-\frac{2}{3\gamma}}{n-\frac{2}{3\gamma}}\frac{na_s}{1-\frac{2}{3\gamma}\left(\frac{n-\frac{2}{3\gamma}}{n-1}\right)^{n-1}}\left(\frac{t}{t_s}\right)^{\frac{2}{3\gamma}} \nonumber
\\&+& \frac{a_s}{\frac{3\gamma}{2}\left(\frac{n-1}{n-\frac{2}{3\gamma}}\right)^{n-1}-1}\left(1-\frac{1-\frac{2}{3\gamma}}{n-\frac{2}{3\gamma}}\frac{t}{t_s}\right)^n~,
\eea
with the admissible values of the parameters: $\gamma >0$ and $n \neq 1$.
Now, let us note that one can write down a barotropic equation of state, using (\ref{rho}) and (\ref{p}), as
\be
p(t) = -\frac{c^2}{3}\left[1+2\frac{\ddot{a}(t)a(t)}{\dot{a}^2(t)}\right] \varrho(t)~,
\ee
and so the effective barotropic index is given by
\be
\label{wt}
w(t) = -\frac{c^2}{3}\left[1+2\frac{\ddot{a}(t)a(t)}{\dot{a}^2(t)}\right] = \frac{c^2}{3}\left[2 q(t) - 1 \right]~,
\ee
where we have introduced the deceleration parameter
\be
\label{q}
q(t) = - \frac{\ddot{a}a}{\dot{a}^2}~.
\ee
\begin{figure}[ht]
 \begin{center}
 \scalebox{0.46}{\includegraphics[angle=0]{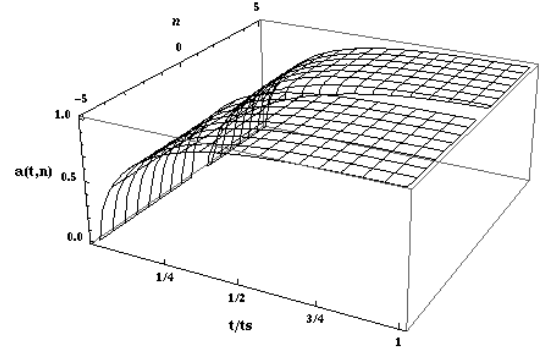}}
 \caption{The scale factor (\ref{SFF}) as a function of time and the parameter $n$ ($n \neq 1$, $a_s=1$, $t_s=10$, $\gamma=1$). We have removed
a piece which corresponds to $n \sim 1$ ($a \sim \infty$ for this value). The Universe starts with a big-bang at $t=0$, where $a=0$, and reaches a $w$-singularity at $t=t_s$, where $a=a_s$.} \label{fig1}
 \end{center}
 \end{figure}
\begin{figure}[ht]
 \begin{center}
 \scalebox{0.46}{\includegraphics[angle=0]{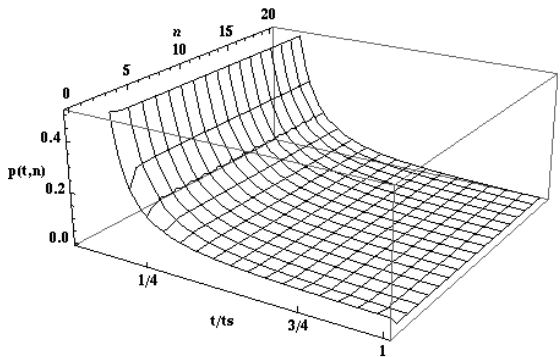}}
\caption{The pressure (\ref{p}) as a function of time and the parameter $n$ ($a_s=1$, $t_s=10$, $\gamma=1$). The Universe begins with a big-bang
at $t=0$, where $p \to \infty$, and reaches a $w$-singularity at $t=t_s$, where $p =0$.}
\label{fig3}
 \end{center}
 \end{figure}
\begin{figure}[ht]
 \begin{center}
 \scalebox{0.46}{\includegraphics[angle=0]{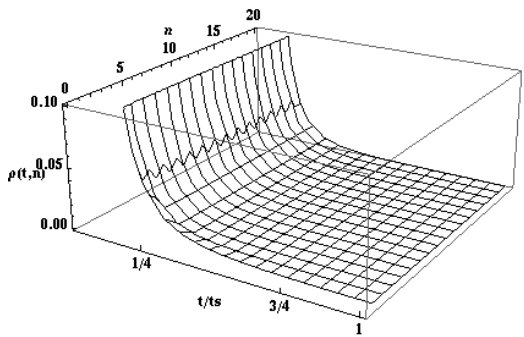}}
\caption{The energy density (\ref{rho}) as a function of time and the parameter $n$ ($a_s=1$, $t_s=10$, $\gamma=1$). The Universe begins with a big-bang at $t=0$, where $\varrho \to \infty$, and reaches a $w$-singularity, where $\varrho =0$.}
\label{fig4}
 \end{center}
 \end{figure}
\begin{figure}[ht]
 \begin{center}
 \scalebox{0.46}{\includegraphics[angle=0]{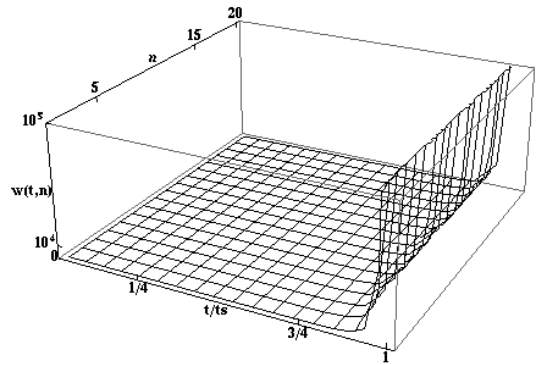}}
\caption{An effective barotropic index (\ref{wt}) as a function of time and the parameter $n$ ($a_s=1$, $t_s=10$, $\gamma=1$). It is zero
at the big-bang and infinity at the $w$-singularity.}
\label{fig2}
 \end{center}
 \end{figure}
On the other hand, the Hubble parameter and its derivatives read as
\bea
H(t) &=& \frac{\dot{a}}{a}~,\\
\dot{H}(t) &=& - 4 \pi G \left(\varrho + \frac{p}{c^2} \right)~,\\
\ddot{H}(t) &=& \frac{\dddot{a}}{a} - 2 \frac{\ddot{a}\dot{a}}{a^2} - \dot{H}H~.
\eea
We now note that at the time $t=t_s$ we have
\bea
H(t_s)&=& 0, \hspace{0.3cm} \dot{H}(t_s) = 0, \\
\label{ddotH}
\ddot{H}(t_s) &=& \frac{\dddot{a}(t_s)}{a_s} = \frac{n}{n-1} \frac{1}{t_s^3}
\frac{\left(1-\frac{2}{3\gamma}\right)^2}{\frac{3\gamma}{2} - \left(\frac{n-\frac{2}{3 \gamma}}{n-1}\right)^{n-1}}~.
\eea
In order to have a blow-up of $\ddot{H}(t_s)$, one would either take $n=1$ (which is a peculiar case), or the denominator of
(\ref{ddotH}) would have to be zero. However, as we have checked both analytically and numerically, it does not happen for any of the admissible values of the constants $\gamma>0$ and $n>0$ ($n \neq 1$). The conclusion is that the model does not admit a singularity of the higher derivatives of the Hubble parameter since $\ddot{H}(t_s) \neq 0$ in (\ref{ddotH}), and so it is not of the type IV
singularity according to the classification of Ref. \cite{nojiri}. On the other hand, despite both $\ddot{a}(t_s)$ and $\dot{a}(t_s)$ vanish
in the limit $t \to t_s$, the deceleration parameter (\ref{q}) blows-up to infinity, i.e.,
\be
\label{qts}
q(t_s) = - \frac{\ddot{a}(t_s)a_s}{\dot{a}^2(t_s)} \to \infty~,
\ee
and consequently, there is a blowup of the effective barotropic index (\ref{wt}), i.e.,
\be
w(t_s) = \frac{c^2}{3}\left[2 q(t_s) - 1 \right] \to \infty~.
\ee
Then, we face a very strange singularity. It has vanishing pressure and energy density, a constant scale factor, but the deceleration parameter and, in particular, a time-dependent barotropic index $w(t)$ are singular. This is why we call it a $w$-singularity.

In Figs. 1-4 we give plots of the scale factor (\ref{SFF}), the pressure (\ref{p}), the energy density (\ref{rho}), and the barotropic index $w = p/\varrho$ for $n_s = 1$, $t_s = 10$, $\gamma = 1$, and $n \neq 1$. It shows that the Universe starts with a big-bang at $t=0$ and reaches a $w-$singularity at $t=t_s$ in which the scale factor (Fig. 1), the pressure (Fig. 1), and the energy density (Fig. 3) are finite, and in which the barotropic index $w = p/\varrho$ blows up to infinity (Fig. 4).

Bearing in mind the relations (\ref{rho}), (\ref{p}) and (\ref{wt}) for the scale factor (\ref{SFF}), we uncover an interesting duality between a big-bang and a $w$-singularity for the models under studies. Namely, the big-bang is characterized by $p \to \infty$, $\varrho \to \infty$ and
$w \to 0$ (close to $t=0$, $a(t) \propto t^{2/3}$ and so $q \propto 1/2$ which from (\ref{wt}) gives $w \to 0$), while the $w$-singularity is characterized by $p \to 0$, $\varrho \to 0$ and $w \to \infty$. This means we deal with duality
between the big-bang and the $w$-singularity in the form
\be
\label{duality}
p_{BB} \leftrightarrow \frac{1}{p_w}, \hspace{0.3cm} \varrho_{BB} \leftrightarrow \frac{1}{\varrho_w}, \hspace{0.3cm} w_{BB} \leftrightarrow \frac{1}{w_w}~~.
\ee


Now, let us check if there is a simple standard cosmology limit of the model. In order to do so, we may define the constant $A$ in (\ref{SF1}) in a similar fashion as it was done in Ref. \cite{sfs07}, i.e.,
\be
\label{Adelta}
A = \delta a_s~, \hspace{0.2cm} {\rm and} \hspace{0.2cm} \delta =
\frac{1}{1 - \frac{3\gamma}{2} \left(\frac{n-1}{n-\frac{2}{3\gamma}}\right)^{n-1}}~.
\ee
Using this, the scale factor (\ref{SFF}) reads as
\bea
\label{SFFdelta}
a(t)&=&a_s \left\{ \delta + n\left( 1 - \delta \right) \frac{1-\frac{2}{3\gamma}}{n-\frac{2}{3\gamma}}
\left(\frac{t}{t_s}\right)^{\frac{2}{3\gamma}} \right. \nonumber \\
 &-& \left. \delta
\left[ 1 -\frac{1-\frac{2}{3\gamma}}{n-\frac{2}{3\gamma}} \frac{t}{t_s} \right]^n \right\} ~.
\eea
From (\ref{SFFdelta}), it seems that the standard flat Friedmann model limit could be obtained under the limit $\delta \to 0$ (provided
($\gamma \neq 2/3$ and $n \neq 2/(3\gamma)$). However, it is not the case, since the limit $\delta \to 0$ corresponds to an infinite value of the denominator of (\ref{Adelta}) which never happens since it is exactly an inverse of the denominator of $\ddot{H}(t_s)$ in (\ref{ddotH}) which is neither zero nor infinity. Finally, let us note that a special $\gamma =0$ dust-filled model which preserves the standard early universe evolution processes is obtained from (\ref{SFF}), provided we choose:
\be
a_s = \frac{3n-2}{n} - \frac{2}{3n} \frac{(3n-2)^n}{(3n-3)^{n-1}}~.
\ee
This gives (\ref{SFF}) as
\bea \label{wsin1}
a(t) &=& \frac{2}{3n} \left(3n -3 \right)^{1-n} \left[ \left(3n -2 - \frac{t}{t_s}\right)^n
- \left(3n-2 \right)^n \right. \nonumber \\
&+& \left. \frac{3n}{2} \left( 3n-3 \right)^{n-1}
\left( \frac{t}{t_s} \right)^{\frac{2}{3}} \right]~.
\eea
The form of (\ref{wsin1}) for some special values of $n$ is:
\bea
a_{n=2}(t) &=& \left(\frac{t}{t_s}\right)^{\frac{2}{3}} + \frac{1}{9} \frac{t}{t_s} \left(\frac{t}{t_s} - 8 \right)~,\\
a_{n=\infty}(t) &=& \left(\frac{t}{t_s}\right)^{\frac{2}{3}} + 2 e^{\frac{1}{3}} \left(e^{\left(1 - \frac{t}{t_s} \right)} - 1 \right)~.
\eea
%
%
%

In conclusion, we have found another type of an exotic singularity which we call
a $w$-singularity. It differs from the type IV singularity of Ref.\cite{nojiri}.
The name comes from the fact that all the quantities like the scale factor, the energy density, the pressure are not singular, while the time dependent barotropic index $w(t)$ in the barotropic form of an equation of state is infinite, i.e., for a certain time $t=t_s$ $a =a_s=$ const., $\varrho(t_s) \to 0$, $p(t_s) \to 0$ and $w(t_s) \to \infty$ with $p(t_s) = w(t_s) \varrho(t_s)$. A $w$-singularity is different from
type IV cosmological singularity in that it does not exhibit
the divergence of the higher derivatives of the Hubble parameter. It is also different from the big-brake, since
it does not fulfill explicitly the anti-Chaplygin gas equation of state $p = K/\varrho$ with $K$=const. In the context of physical theories $w$-singularities may appear in $f(R)$ gravity \cite{starob80}, in scalar field models \cite{saridakis}, and in brane cosmologies \cite{yuri}.

The $w$-singularities do not seem to violate the energy conditions:
null $\varrho c^2 + p \geq 0$), weak ($\varrho c^2 \geq 0$ and $\varrho c^2 + p \geq 0$),
strong ($\varrho c^2 + p \geq 0$ and $\varrho c^2 + 3p \geq 0$), and
dominant energy ($\varrho c^2 \geq 0$, $-\varrho c^2 \leq p \leq
\varrho c^2$) since the energy density and the pressure are zero at these singularities.
Of course they reality of $w$-singularities should be verified observationally.

Finally, we have found an interesting duality (\ref{duality}) between the big-bang and the $w$-singularity which refers to
the pressure, the energy density and the barotropic $w$-index.

\section{Acknowledgements}

We acknowledge the support of the Polish Ministry of
Science and Higher Education grant No N N202 1912 34 (years 2008-10).
We thank A.A. Starobinsky, E.N. Saridakis, and V. Sahni for drawing our attention to
Refs. \cite{starob80,saridakis,yuri}.

\end{document}